\documentstyle[12pt]{article}

\begin{document}

\renewcommand{\thefootnote}{\fnsymbol{footnote}}

\begin{center}
{\bf
Seeing the QCD phase transition with phi mesons}
\end{center}

\vspace{0.5cm}
\begin{center}
{M. Asakawa\footnote{Electronic address: yuki@nsdssd.lbl.gov}
\footnote{Address after October 1, 1993: Lawrence Berkeley Laboratory,
Nuclear Science Division, MS 70A-3307, 1 Cyclotron Road, Berkeley,
CA 94720, U.S.A.} and
C. M. Ko\footnote{Electronic address: ko@comp.tamu.edu}
\\}
{\it
Cyclotron Institute and Physics Department\\
Texas A{\&}M University \\
College Station, Texas 77843\\}
\end{center}

\vspace{1cm}
\centerline{\bf Abstract}
\medskip

A double phi peak structure in the dilepton invariant
mass spectrum from ultrarelativistic heavy ion collisions is proposed
as a signal for the phase transition from the quark-gluon plasma
to the hadronic matter. The low mass phi peak results from
the decay of phi mesons with reduced in-medium mass during the transition.
Furthermore, the measurement of the transverse momentum
distribution of these low mass phi mesons offers a viable means
for determining the temperature of the phase transition.

\vspace{0.5cm}

\vspace{3cm}
\centerline{\it Physics Letters B in press}
\newpage

Heavy-ion experiments offer the possibility to
create in the laboratory the deconfined quark-gluon plasma.
This allows us thus the opportunity to study both
the properties of the quark-gluon plasma and the nature of
its transition to the hadronic matter.
Many experimental observables have been proposed as signatures for
its existence \cite{muqm91}.  Since the quark-gluon
plasma exists only at the initial stage of heavy-ion
collisions, electromagnetic probes such as lepton pairs and photons
are therefore more suitable than  hadronic probes as they are
not affected by final-state interactions.  Indeed,
we have shown recently that the $M_T$ scaling in the dilepton spectrum
is a plausible signature for the quark-gluon plasma expected to be
formed in
future experiments at the Relativistic Heavy Ion Collider (RHIC)
and the Large Hadron Collider (LHC) \cite{ak93}.

In the present paper, we would like to propose
a signature for
identifying the phase transition
of the quark-gluon plasma to the hadronic matter in ultrarelativistic
heavy ion collisions.
In  hot hadronic matter the phi meson mass is expected to decrease
as a result of  the partial restoration of
chiral symmetry \cite{pi82,ha85,ha92}.
If a first-order phase transition
between the quark-gluon plasma and the hadronic matter occurs in
heavy ion collisions, then a low mass phi peak besides the normal one
will appear in the dilepton spectrum.  This is due to the
nonnegligible duration time for the system to
stay near the transition  temperature (about 10 fm/c in boost invariant
hydrodynamical calculations with transverse flow)
compared with the lifetime
of a phi meson in the vacuum ($\sim 45$ fm), so
the contribution to dileptons
from phi meson decays in the mixed phase becomes comparable
to that from phi meson decays at freezeout.

This scenario can be described more quantitatively.
We first determine the phi meson mass at finite temperatures
using the QCD sum rules \cite{svz},
which relate via the dispersion relation the phi meson
mass to the quark and gluon condensates in the matter.
Following the treatment of Ref. \cite{hl92}, we include
up to dimension 6 all
scalar operators and tensor operators with twist 2.
The temperature effect is then included via the change of
the condensates.  For example, the strange quark
condensate at finite temperatures $\langle\bar{s}s\rangle_T$ is
approximately related
to its value at zero temperature $\langle\bar{s}s\rangle_0$,
\begin{equation}\label{sbars}
\langle \bar{s}s \rangle _T \approx
\langle \bar{s}s \rangle _0 +
\sum_{h} \langle \bar{s}s \rangle _h \rho_h,
\end{equation}
where $\langle\bar{s}s\rangle_h$ and $\rho_h$ are, respectively, the
strangeness content and density of hadron $h$.
Assuming that all hadron densities are given by their equilibrium values,
the resulting temperature dependence of the strangeness condensate
can be calculated.

We show in Fig. 1 the temperature dependence of the phi meson mass
at rest $m_\phi$
in a hot hadronic matter with zero baryon density.
We see that the phi meson mass decreases at high temperatures.
The decrease of phi meson mass is mainly
due to the presence of strange particles, which have larger
strangeness content than nonstrange particles,
in the  hot matter.  Because of the
relatively small number of strange particles in the hot matter,
the reduction in phi meson mass is less than that for the rho meson.
Studies based on QCD sum rules \cite{fur90} show that
the temperature
dependence of the rho meson mass is approximately given by
\begin{equation}\label{rho}
\frac{m_{\rho} (T)}{ m_{\rho} (T=0)}\approx
\left [1-\left (\frac{T}{T_c}\right )^2\right ]^{1/6},
\end{equation}
where $T_c$ is the critical temperature for the chiral restoration
transition.  Recent lattice calculations have shown that this
temperature is similar to that for the quark-gluon plasma
to hadronic matter transition \cite{kar90}.  However, the omega meson mass
does not change much with the temperature as it has
a different isospin structure from that of the rho meson \cite{hkl92}.
We shall thus in the following assume that the omega meson mass
is independent of the temperature.
Details of the calculation
for the temperature dependence of the phi meson mass
will be reported elsewhere \cite{ak92}.

We note that calculations based on effective hadronic Lagrangians often
give rather different behavior for  the vector meson masses at
finite temperatures \cite{gk91,song93}.  Also,  some QCD
sum rule studies \cite{dey90} lead to results that are at variance with
those of  Refs. \cite{hl92,fur90}.
However, recent lattice QCD calculations \cite{boyd93}
show that both rho and phi masses are reduced at high temperatures.
Our dropping phi meson mass at finite temperatures
is thus likely to be correct.

To calculate the dilepton yield from the hot hadronic matter,
we assume that in ultrarelativistic heavy
ion collisions the system has a
cylindrical symmetry and is in thermal equilibrium.
At temperature $T$, the number of phi mesons that decay into lepton
pairs per unit time and unit phase space is given by
\begin{equation}\label{dec1}
dN_{\bar{\ell}\ell} = \frac{g_\phi}{(2\pi)^3}\cdot\frac{1}{\gamma _\phi}
\cdot \Gamma_{\bar{\ell}{\ell}}(T)
e^{-\frac{p\cdot u}{T}}d^4 x d^3 p,
\end{equation}
where $g_\phi = 3$ is the degeneracy of the phi meson and $u$ is the
four-dimensional flow velocity.
The temperature-dependent decay width of the phi meson into a lepton
pair is denoted by $\Gamma_{\bar{\ell}\ell}(T)$ and
is proportional to the phi meson mass.
The effect of the Lorentz dilatation is
taken into account through the Lorentz factor of the phi meson $\gamma_\phi$.
We note that in Eq. (\ref{dec1}) we have used the Boltzmann
distribution for phi mesons.
Assuming boost-invariance, we can integrate
Eq. (\ref{dec1}) over the phase space and obtain
\begin{equation}\label{dec2}
\frac{dN_{\bar{\ell}{\ell}}}{dMdy}= \frac{g_\phi}{\pi}\int
\! f_h (\rho, \tau) \, T m^2 _\phi(T)
\Gamma_{\bar{\ell}\ell}(T) F_{\phi}(M, m_\phi (T) )
K_1 \left ( \frac{m_\phi (T)}{T} \right ) \tau \rho d\tau d\rho,
\end{equation}
where $M$ and $y$ are the invariant mass and rapidity of the lepton pair,
respectively.  In the above, we have denoted by
$\tau$ the proper time, $\rho$ the radial coordinate,
$K_1$ the modified Bessel function of the second kind, and $f_h$
the volume fraction of the hadron phase. In Eq. (\ref{dec2}) there is no
dependence on the transverse expansion as
the effects of the Lorentz dilatation and Lorentz contraction
cancel each other.

The total width of a phi meson in hot hadronic matter is expected
to change as well.   Shuryak {\it et al.} \cite{ls91,sh92} have found that
if the phi meson mass is assumed to be unchanged at finite temperatures
its width is then approximately doubled as a result of the
attractive kaon  potential.   With the phi meson mass reduced to much below
twice kaon mass at high temperatures as shown in Fig. 1,
this effect will not be important  in our studies.
However, there will be a collisional broadening of the phi meson
width due to its interaction with  pions.  Bi and Rafelski \cite{bi91} have
estimated that this would also double the phi meson width.
A more detailed study of the phi meson width in a hot hadronic matter
has been carried out in Ref. \cite{sei93}.  Taking into account the
scattering of a phi meson with pions, kaons, rho mesons, and phi mesons
in the matter, it is found that the width of a phi meson is less than
10 MeV at all temperatures.  Also, the experimental mass resolution
in future RHIC dilepton measurements
is about $5\sim 10$ MeV around the phi meson mass \cite{ph92}
and is comparable to the  phi meson width discussed in the above,
we have thus introduced in Eq. (\ref{dec2})
a normalized smearing function $F_{\phi} (M, m_\phi (T) )$ of the
Gaussian form
\begin{equation}
F_{\phi} (M, m_\phi (T)) = \frac{1}{\sigma\sqrt{2\pi}} e^{-(M-m_\phi (T))^2
/ 2\sigma^2 },
\end{equation}
where $\sigma$ is a constant and is taken to be $\sigma=10\,{\rm MeV}$.

We also include dileptons from phi meson decays
at  freezeout.  For a transversely uniform system with a
freezeout hypersurface $\tau=\tau_f$, this contribution is given by
\begin{equation}\label{dec3}
\frac{dN_{\bar{\ell}\ell}}{dMdy} = \frac{g_\phi}{2\pi}r_f^2 \tau_f T_f
m^2_{\phi} B_{\bar{\ell}{\ell}}F_{\phi}(M, m_\phi)
K_2 \left (\frac{m_\phi}{T_f}\right ),
\end{equation}
where $r_f$ is the radius of the system at freezeout,
$B_{\bar{\ell}{\ell}}$ is the branching ratio of a phi meson decaying
into a lepton pair, and $K_2$ is the modified Bessel function of
the second kind.  The phi meson mass at freezeout $m_\phi$
is taken to be its free mass.

The contribution to dileptons from omega mesons in both the hot phase
and at freezeout can be similarly evaluated.  On the other hand, we
include the rho meson contribution to dileptons via the
pion-pion annihilation.  This should be similar to treating directly
the rho decay into dilepton.

Using the hydrodynamical code of Ref. \cite{lm91},
we have carried out a boost-invariant hydrodynamical calculation
with transverse flow for a hot system that is expected to be formed in
the central collisions of ${\rm ^{197}\!Au+^{197}\!Au}$ at RHIC and LHC.
For the equation of state, we take both the quark-gluon
plasma and the pionic matter as free gases.  The phase transition
between the two phases is then of the first order.
The initial radial velocity
at the surface of the cylinder is chosen to be $v_0 = 0$.
The critical temperature $T_c=180$ MeV and the
freezeout temperature $T_f=120$
MeV are taken to be the same as in our previous work \cite{ak93}.
We include dilepton production from the phi and omega decays,
the $\pi\pi$ annihilation, and the $q\bar q$ annihilation in
the quark-gluon plasma.
For dileptons from the $\pi\pi$ and $q\bar q$ annihilations,
we do not introduce the smearing function as the yield does not
change much within the experimental resolution.

In Fig. 2 the dilepton spectrum in the rho, omega,
and phi region is shown by the solid curve
for the standard initial temperature $T_i=250\,{\rm MeV}$
and initial proper time $\tau_0 = 1\,{\rm fm}$.  We indeed see
a second phi peak around 880 MeV between the omega meson
and the normal phi meson.  We note that it
is exclusively from phi meson decays in the mixed phase.  Due to
its temperature-dependent mass, the rho meson peak not only shifts
to lower masses but is also much broadened.
Also shown in Fig. 2 by the dotted curve is the result
from the hadronic scenario
in which the initial state is taken to be a hot hadronic matter
at a temperature just below $T_c$.
In this case, the dropping phi meson mass only leads to a
slight enhancement of
the low mass side of the phi meson peak.  Because of
the shorter lifetime of the hot phase in the
hadronic scenario, the dilepton yield is also seen to be substantially
reduced.

Our results remain essentially unchanged if we
use the same critical temperature $T_c=180$ MeV but different
initial temperatures.  In particular, we still see a double
phi peak structure using
an initial temperature of 450 MeV as in the hot glue scenario
that has been suggested by recent studies based on parton scatterings
\cite{geiger92,shu92,km92}.

Since the transverse expansion velocity during the mixed phase is
relatively small, the low mass phi mesons should also provide
information about the critical temperature of the phase transition.
This temperature can be extracted from the transverse momentum
distribution of phi mesons corresponding to the low mass peak.
In Fig. 3, we show the slope parameter of the dilepton distribution at
small transverse momenta as a function of the initial temperature.
The initial proper time is taken to be
$\tau_0=\,1$ fm, independent of the initial temperature.
The solid curve is the slope parameter of the low mass peak at about
880 MeV.  It depends weakly on the initial temperature and is close to
the critical temperature $T_c$ assumed in our study and shown by the
dashed line.  The dotted curve is the slope parameter of the normal
phi meson peak at 1019 MeV.  It is much larger than the slope
parameter of the low mass peak and increases with the initial
temperature as a result of the appreciable transverse flow at
freezeout.

The determination of the critical temperature for the quark-gluon
plasma to hadronic matter transition from the transverse momentum
distribution of the rho meson has
been proposed by Seibert \cite{sei92}.  As dileptons
from rho meson decays during the expansion of the hadronic matter
are not negligible, its slope parameter does not give as accurate
a measurement of the transition temperature as the low mass phi mesons
proposed in the present study.

In our study, we have assumed that the rho meson mass vanishes
at the critical temperature as shown in Eq. (\ref{rho}), and there is
thus no dileptons from rho decays in the mixed phase.
If the rho meson mass remains finite at this temperature, then there
will also be a low mass rho peak in the dilepton spectrum.
However, to observe it will be difficult
as its invariant mass is quite low and there is also a
large dilepton background from the Dalitz decays.

The double peak structure has also been suggested in the literature
for the $J/\psi$ due to  possible
changes of the string tension in hot matter
\cite{mi86}.  However, this turns out not to be the case as already
pointed out in Ref. \cite{hi91}.  Since the
lifetime of $J/\psi$ is much longer than the duration time of the
hot phase expected in ultrarelativistic heavy ion collisions, only a
negligible number of $J/\psi$ decay in the hot phase and the low mass
$J/\psi$ peak can thus not be seen.

In our calculation, we have used
the normal Maxwell construction to determine
the volume fraction of the quark-gluon plasma and the
hadronic matter in the mixed phase. Including the effect of
supercooling \cite{ck92}
would increase the duration time of the mixed phase
and make the double peak structure of the phi meson more prominent.

On the other hand, the second low mass peak may become less visible
because of the following reasons.
1) If the largest fraction of the drop in
entropy density in a first-order phase transition occurs
already above $T_c$, where the system is still deconfined and no bound
phi state exists, so the mixed phase lives much shorter.
2) The phi meson width
in the mixed phase may be broadened by its interactions with
partons in the quark-gluon plasma.  3) If the phase
transition between the quark-gluon plasma and the hadronic
matter is not of the first order as assumed in the above calculation
but is of a second order or a crossover.  To make definite statements
in these cases requires, however, further detailed studies.

In conclusion, we have shown that the recent predictions
from the QCD sum rules on the change of meson masses
in hot hadronic matter have dramatic effects on the
dilepton invariant mass spectrum
from ultrarelativistic heavy ion collisions.
Due to the dropping phi meson mass in a hot matter,
a distinct low mass peak besides the normal one appears in the
dilepton spectrum
if the quark-gluon plasma to hadronic matter phase transition
in heavy ion collisions is a strong first-order one.
This low mass phi meson peak is thus a viable signal for the phase
transition from the quark-gluon plasma to the hadronic matter.
Furthermore, the measurement of the transverse momentum
distribution of these low mass phi mesons
also allows us to determine the transition temperature
between these two phases of matter.

\bigskip

This work was supported in part by the National Science Foundation under
Grant No. PHY-9212209 and the Welch Foundation under
Grant No. A-1110.

\bigskip
\centerline{\bf Figure Captions}
\vskip 15pt
\begin{description}
\item[Fig. 1] The temperature-dependent phi meson mass in a hot
hadronic matter.
\item[Fig. 2] The dilepton invariant mass spectrum at the central rapidity.
The solid curve is the result from the hydrodynamical calculations with the
initial temperature $T_i=250$ MeV and proper time $\tau_0=1$ fm.
The dotted curve is obtained from the hadronic scenario assuming that
the initial phase is a hadronic matter.
\item[Fig. 3] The slope parameter of the phi meson transverse momentum
distribution as a function of the initial temperature.
Solid and dotted curves correspond to the low mass
peak and the normal peak, respectively.  The dashed line is the critical
temperature for the quark-gluon to hadronic matter transition.
\end{description}

\end{document}